# A scalable compact model for the static drain current of graphene FETs

Nikolaos Mavredakis, Anibal Pacheco-Sanchez, Oihana Txoperena, Elias Torres, and David Jiménez

*Abstract*—The main target of this article is to propose for the first time a physics-based continuous and symmetric compact model that accurately captures *IV* experimental dependencies induced by geometrical scaling effects for graphene transistor (GFET) technologies. Such a scalable model is an indispensable ingredient for the boost of large-scale GFET applications, as it has been already proved in solid industry-based CMOS technologies. Dependencies of the physical model parameters on channel dimensions, are thoroughly investigated, and semi-empirical expressions are derived, which precisely characterize such behaviors for an industry-based GFET technology, as well as for others developed in research labs. This work aims at the establishment of the first industry standard GFET compact model that can be integrated in circuit simulation tools and hence, can contribute to the update of GFET technology from the research level to massive industry production.

## I. INTRODUCTION

NOWADAYS, classical (constant-field) metal-oxide-semiconductor field-effect transistors' (MOSFETs) scaling [1, §3.3], [2, §5.15.2] is no longer valid. As MOSFETs approach sub-nanometer size, limitations due to significant oxide leakages and short-channel effects, have led to the rise of emergent devices such as graphene field-effect transistors (GFETs) suitable for analog/RF applications [3], and other 2D material-based FETs appropriate for digital circuitry [4]. Recently, we have proposed a physics-based modular GFET compact model, proper for GFET-based integrated circuits (ICs) [5] (Note that a compact model for other 2D-FETs is introduced in [4]). To establish such a model as the first industry standard, scalability features are essential. In more detail, reliable compact models suitable for large-scale ICs, should accurately describe both the underlying physics and the device footprint (channel width $W$ and length $L$) scaling of all devices within a chip, with one parameter set.

As a rule, compact modeling engineers derive firstly the basic parameters for the wide-long channel devices within a process, where a one-dimensional (1D) analysis is applied as, according to the gradual channel approximation, vertical field is much stronger than longitudinal and lateral fields along a microscopic slice of the channel in such large FETs [1, §5.1], [2, §4.1], [6, §3.1]. To the contrary, in small devices, source and drain can affect the channel similarly as the gate, due to their proximity to all channel points [2, §5.4]. Thus, 2D or 3D analysis is required to deal with short length and/or narrow width channel physical mechanisms, which can be implemented numerically in a TCAD simulator. Such approaches, although accurate [7], are not appropriate for fast calculations in circuit simulators. Hence, semi-empirical approximations have been considered for scalable compact models, where the arduous multidimensional effects are fragmentized into simplified uncorrelated phenomena investigated individually. According to the latter methodology, the basic *IV* principles of the wide-long 1D analysis are adjusted appropriately to apply in short and/or narrow channels [2, §5.4]. Such semi-empirical content can be found in all standard CMOS compact models, where linear and/or quadratic, as well as exponential geometrical relations, are implemented [1, §6.7], [2, §5.7], [8]-[12]. The geometrical scaling of basic model parameters such as threshold voltage $V_{TH}$ [1, §5.3], [2, §5.4], [9], drain(source) series resistance $R_{d(s)}$ [1, §3.6.1], [2, §6.8.2], [6, §10.2.1] and low-field mobility $\mu$ [1, §6.6.1], [8], [13], [14] have been thoroughly investigated and modelled in MOSFETs.

In this work, such critical scaling effects are investigated, for an industry-based GFET technology (Graphenea PF1) [15], as well as for different published GFET technologies, mainly developed in research labs [16]-[20]. Geometrical variations have been earlier experimentally studied for Dirac voltage $V_{Dirac}$ ($V_{TH}$ equivalent in GFETs) in [17], [18], for $R_{d(s)}$ in [20]-[23], for $\mu$ in [17], [23]-[27] and for impurities-related residual charge $n_{res}$ in [26], however, such

This work has received funding from the European Union's Horizon 2020 research and innovation programme under grant agreements No GrapheneCore3 881603, from Ministerio de Ciencia, Innovación y Universidades under grant agreements RTI2018-097876-B-C21(MCIU/AEI/FEDER, UE), FJC2020-046213-I and PID2021-127840NB-I00 (MCIN/AEI/FEDER, UE) and by the European Union Regional Development Fund within the framework of the ERDF Operational Program of Catalonia 2014-2020 with the support of the Department de Recerca i Universitat, with a grant of 50% of total cost eligible. GraphCAT project reference: 001-P-001702.

N. Mavredakis, A. Pacheco and D. Jiménez are with the Departament d'Enginyeria Electrònica, Escola d'Enginyeria, Universitat Autònoma de Barcelona, Bellaterra 08193, Spain. (e-mail: Nikolaos.mavredakis@uab.cat).
Oihana Txoperena and Elias Torres are with Graphenea Semiconductor SLU., Paseo Mikeletegi 83, 20009 San Sebastián, Spain

parameters' scaling observations are still missing in circuit-friendly models. In this work, the compact model [5, §2.1] has been updated accordingly to account for such dependencies for the first time, for all GFET technologies under test, by introducing new scaling parameters and equations. Such scaling features are not included in previous models describing GFETs with different lengths [23], [25], [28]. For the extraction of the basic parameters of the model ($V_{Dirac}$-related flat-band voltage $V_{G0}$, $R_{d(s)}$, $\mu$, intrinsic mobility degradation $\theta_{int}$ [29] and $n_{res}$-related parameter $\Delta$) [5] several methodologies have been proposed [29], [30] where the one presented in [29], is applied here. $IV$ hysteresis effects are also modelled, provided that trap-affected and trap-reduced measurement setups are available [31]. Finally, detailed derivations are included in the model to account separately for $p$- and $n$-type regions of operation [32], while the model is accurately benchmarked with the industry standard Gummel Symmetry Test (GST), establishing its symmetry and continuity [2, §K.1], [33].

## II. DEVICES AND MEASUREMENTS

Six back-gated GFETs from Graphenea PF1 technology ($W/L$=50/60, 50/70, 50/100, 10/50, 40/50, 100/50 $\mu m/\mu m$) have been measured with both trap-affected (forward-backward sweeps) and trap-reduced (opposing pulses) $IV$ setups [31], to validate the proposed geometrical scaling module of the model. Several drain $V_{DS}$ (0.05, 0.1, 0.3, 0.5, 0.7, 1 V) and gate $V_G$ (0 to 50 V with 0.5 V step) voltages, have been selected to embrace all operating regimes. A 90 nm thick $SiO_2$ is used as a back-gate oxide [15]. The device structure (cf. Fig. 1), due to its simple structure and low processing step number, diminishes variability-reliability issues; such devices are optimal for sensing and bioelectronics applications. The model is also tested for trap-reduced $IV$ experiments extracted from large GFETs of [20]. The aforementioned Graphenea PF1 and [20] technologies, include only GFETs with large $W$, $L$, hence, additional experimental data from short-channel devices taken from literature [16]-[19], are used for an optimum validation of the model. To our knowledge, $IV$ experiments for different $W$ values within a specific GFET technology are not available (apart from Graphenea GFETs examined here).

## III. MODEL DERIVATIONS

The drain current $I_D$ model in [5, §2.1], is here updated as:

$$I_D = \frac{W\mu_{ueffp(n)}\int_{V_S}^{V_d}|Q_{gr}|dV}{L} = \frac{W\mu_{up(n)}\int_{V_S}^{V_d}|Q_{gr}|dV}{L_{effp(n)}} \quad (1)$$

so as to enable distinct $\mu_{p(n)}$ [32] and $R_{dsp(n)}$ ($R_s=R_d=R_{ds}$ here) models for $p$- and $n$-type conduction regions, respectively (see Appendix for explicit derivations); $R_s=R_d$ are placed between internal and external source and drain nodes in the Verilog-A model code, respectively, and thus, $I_D$ decrease due to $R_{ds}$, is correctly calculated. $Q_{gr}$ is the total graphene charge [5, §2.1] while $\mu_{p(n)}$ reduction due to $\theta_{int}$ [29], is included in $\mu_{up(n)}$ terms. Velocity saturation (VS) induced effective mobility $\mu_{ueffp(n)}$, and consequently effective length

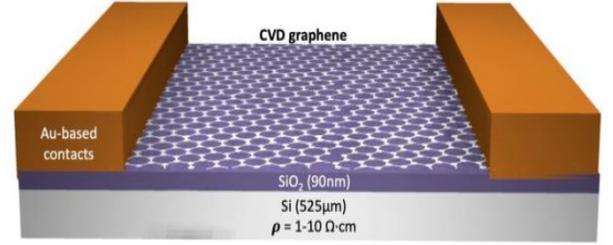

**Fig.1**. Graphenea GFET schematic cross-section not drawn to scale. Graphene under metal contacts is not shown.

$L_{effp(n)}$, are given by [5, §2.1]:

$$\mu_{ueffp(n)} = \frac{\mu_{up(n)}}{\left[1+\left(\frac{\mu_{up(n)}V_{DS}}{u_{sat}L}\right)^b\right]^{1/b}} \xrightarrow{eq. 1} L_{effp(n)} = L\frac{\mu_{up(n)}}{\mu_{ueffp(n)}} \quad (2)$$

where $u_{sat}$ is the saturation velocity and $b$ a VS smoothing parameter [5, §2.1]. In the following part, geometrical effects of the main GFET parameters will be thoroughly characterized and properly modeled, after highlighting their relation with the corresponding scaling effects in MOSFETs.

It has been experimentally shown in CMOS technologies that the shrinkage of device $L$ and $W$, leads to a variation of $V_{TH}$ from its wide-long channel value [1, §5.3], [9]. This can be justified by small dimension phenomena where the effect of source and drain on channel charge increases as $W$, $L$ decrease for a specific $V_G$ point, causing significant modifications on $I_D$ (charge sharing). This mechanism has been modeled in MOSFETs by adopting an effective threshold voltage approximation. [2, §5.4], [1, §5.3], [9]. In more detail, to accurately capture geometrical $V_{TH}$ behavior, industry standard CMOS models (BSIM-Bulk [10], PSP [12], EKV3 [11]) employ scaling equations including 1st- and 2nd-order together with exponential and logarithmic terms [10, eq. 1.18], [11, eq. 2.32, 2.35], [12, eq. 3.18-3.20].

As in CMOS, the variation of measured $V_{Dirac}$ in short-channel GFETs can be associated with an increased contribution of source and drain potentials to the channel potential [17], [18]. For instance, for short-channel GFETs in $p$-type conduction, a more positive voltage is required to drop to minimum $I_D$, and hence $V_{Dirac}$ increases similarly to the $V_{TH}$ in standard $p$-MOSFETs. [17]. Although experimental confirmation is missing, increased edge effects in narrow ($W$ of tens of nm or below) graphene nanoribbon transistors can contribute to the increase of $V_{TH}$ [35] but such devices are out of the scope of the present study. Shifts in $V_{Dirac}$ have also been noticed even for wide-long devices, which can be attributed to the polycrystalline nature of graphene [20]. The following 2nd-order equation is formulated for the geometrical scaling of $V_{G0}$:

$$V_{G0} = V_{GS0} \cdot \left(1 + \frac{V_{GS0W}}{W} + \frac{V_{GS0W2}}{W^2} + \frac{V_{GS0L}}{L} + \frac{V_{GS0L2}}{L^2} + \frac{V_{GS0WL}}{WL}\right) \quad (3)$$

where wide-long channel ($V_{GS0}$), $W$ ($V_{GS0W}$, $V_{GS0W2}$), $L$ ($V_{GS0L}$, $V_{GS0L2}$) and $WL$ ($V_{GS0WL}$) scaling parameters are defined, respectively, as in CMOS models [10]-[12].

Considering $R_{ds}$, it does not scale with $L$ thus it can no longer be neglected in short channels where channel resistance declines. In such case, $R_{ds}$ can decrease the transconductance $g_m$ and thus, critically deteriorate the performance of MOSFETs [1, §3.6.1]. $R_{ds}$ reads as the sum of contact ($R_c$), main sheet ($R_{sh}$) and extrinsic sheet ($R_{shext}$) resistances, respectively; $R_{sh}+R_{shext}$ is inversely proportional to $W$ while $R_{shext}$ can also be bias dependent, as it lies in the overlap region between the gate and the contact [2, §6.8.2], [6, §10.2.1]. BSIM-Bulk, PSP standard CMOS models predict a constant diffusion $R_c$ term [10, eq. 1.159-1.161], [12, eq. 3.302-3.303] as in [2, eq. 6.8.4], while a first-order $\sim 1/W$ geometrical scaling relation is derived for $R_{sh}+R_{shext}$, (including a bias dependency for $R_{shext}$) which is further tuned with exponential [10, eq. 1.159-1.161] or 2$^{nd}$-order parameters [11, eq. 2.297-2.300], [12, eq. 3.67].

In GFETs, $R_{ds}$ can also split into two components in a metal-graphene interface; $R_c$ between the metal contact and the coated-graphene region and $R_{sh}$ between the coated and the uncoated-graphene region [20]-[23]. Note that $R_c$, which remains high in GFETs, can be reduced by using palladium contacts [34]. Alike to CMOS, $R_{ds}$ is significant in short-channel devices, and this can justify reported asymmetries between $p$-and $n$-type regions in short GFETs, as different parasitic diodes ($p$-$n$ at $p$-type and $n$-$n$ at $n$-type regime, respectively) can arise between the graphene channel and the graphene under the contact, inducing uneven $R_{ds}$ ($R_{dsp}$, $R_{dsn}$, respectively) at the two operation regions [18], [21]. $R_{ds}$ geometrical effects are described in our model as:

$$R_{dsp(n)} = R_{ds0p(n)} + R_{coeffp(n)} \cdot \left(\frac{1}{W}\right)^{R_{expp(n)}} \quad (4)$$

where $R_{ds0}$ is the constant term and the $\sim 1/W$ trend of $R_{sh}$ is described by $R_{coeff}$ coefficient while $R_{exp}$ exponent is also employed for optimum model calibration.

Regarding $\mu$, CMOS experimental data have revealed a proportional to $W$, $L$ dependence, which is precisely described by semi-empirical models [1, eq. 6.156], [8, eq. 8]. Short-channel $\mu$ lessening might be caused by the non-uniform distribution of inversion charge across the channel and/or by long-range Coulomb scattering effects [13] while the diminishing of $\mu$ in narrow devices, is probably due to a degradation of the Si crystal in the thin edges [14]. $W$, $L$, $WL$ scaling expressions have been used for $\mu$ in CMOS models [10, eq. 1.22], [11, eq. 2.37-2.38], [12, eq. 3.50-3.54].

In GFETs, peak $g_m$ has been recorded to degrade due to a $\mu$ reduction at short channels [17]. Such behavior can be induced either by an e-beam damage during metal-contact fabrication process [24], by contact-metal doping effects [27] and/or by a strong quasi-ballistic transport which becomes more critical as $L$ decreases [23]-[26]. Besides, long-range Coulomb scattering effects can also decrease $\mu$ in GFETs [26], similarly to CMOS. The proposed $\mu$ model follows a 2$^{nd}$-order relation as:

$$\mu_{p(n)} = \mu_{0p(n)} \cdot \left(1 + \frac{\mu_{Lp(n)}}{L} + \frac{\mu_{L2p(n)}}{L^2}\right) \quad (5)$$

where wide-long channel ($\mu_{0p(n)}$) and $L$ scaling parameters ($\mu_L$, $\mu_{L2}$) are extracted. The latter projects a $\mu$ reduction for short $L$ ($nm$ range). $n_{res}$ has been shown to scale inversely with $\mu$ in terms of $L$ [26, Fig. 5], and thus a similar 2$^{nd}$-order expression is derived (as in $\mu$) for $\Delta$ parameter:

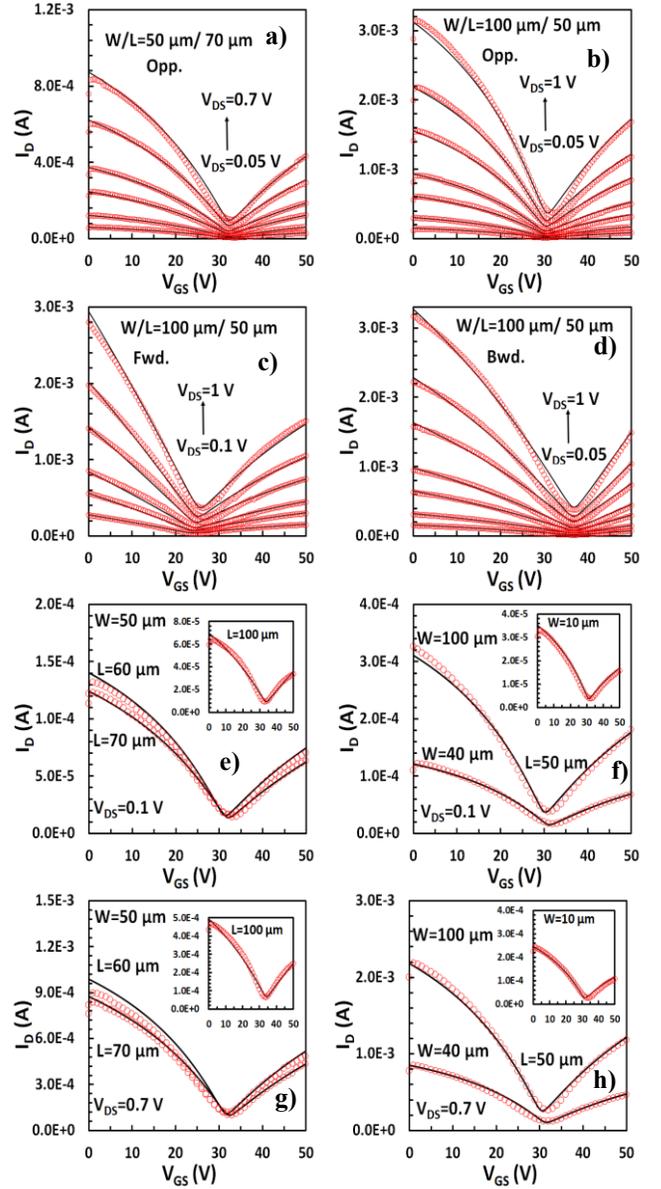

**Fig.2**. Transfer characteristics for Graphenea GFETs with a) width $W=50$ $\mu m$ and length $L=70$ $\mu m$, b-d) $W=100$ $\mu m$ and $L=50$ $\mu m$, e, g) $W=50$ $\mu m$ and $L=60, 70$ $\mu m$ (inset: $L=100$ $\mu m$) and f, h) $L=50$ $\mu m$ and $W=40, 100$ $\mu m$ (inset: $W=10$ $\mu m$) at $V_{DS}=0.05, 0.1, 0.2, 0.3, 0.5, 0.7, 1$ $V$ (a-d), $0.1$ $V$ (e, f) and $0.7$ $V$ (g, h). Opposing pulse sweeps: a, b, e-h, staircase sweeps: c (forward), d (backward). Markers: measurements, lines: model.

$$\Delta = \text{delta}\left(1 + \frac{\text{deltaL}}{L} + \frac{\text{deltaL2}}{L^2}\right) \quad (6)$$

where wide-long channel (*delta*) and $L$ scaling parameters (*deltaL*, *deltaL2*) are defined. The latter predicts an increase of $\Delta$ (and $n_{res}$), as $L$ gets significantly short, due to the inversely proportional relation of $\Delta$ with $\mu$. $W$ scaling of $\mu$, $\Delta$ parameters at narrow GFETs has not been studied yet due to lack of experiments; Graphenea PF1 GFETs with different $W$ ($>10$ $\mu m$ for every case) investigated here, do not seem to introduce any $W$ dependencies to $\mu$, $\Delta$.

IV. RESULTS – DISCUSSION

The basic parameters ($V_{G0}$, $R_{ds}$, $\mu$, $\theta_{int}$, $\Delta$) are extracted from experimental data of each GFET under test according to [29] at low $V_{DS}$, while VS-related $u_{sat}$ and $b$ parameters can be derived from high $V_{DS}$ regime especially at shorter channel lengths [16]; $R_{ds}$, $\mu$ are calculated separately for *p*- and *n*-type regions to consider asymmetries. Geometrical scaling parameters (cf. Table I) are estimated by fitting (3), (5)-(6) with the extracted $V_{G0}$, $\mu_{p(n)}$, $\Delta$ vs. *L* for all GFETs under test, and (3), (4) with the extracted $V_{G0}$, $R_{dsp(n)}$ vs. *W* for Graphenea GFETs. 2$^{nd}$-order parameters dominate at the shorter and 1$^{st}$-order ones at moderate *L*, respectively, while optimization routines are afterwards engaged to better calibrate the calculated parameters after several steps. It is ideal to extract the geometrical scaling parameters from trap-reduced experimental data to ensure identical measurement conditions for every GFET within the same technology. Such experimental procedure is not mentioned in [16]-[19] thus, $K_{tr}$ parameter [31], which describes the hysteresis-induced $V_{Dirac}$ shift at high $V_{DS}$ regime, is extracted for improving the model accuracy in [16], [18], [19] (cf. Table I).

TABLE I
EXTRACTED PARAMETERS

| Parameter | Units | [16] | [17] | [18] | [19] | [20] |
|---|---|---|---|---|---|---|
| $V_{GS0}$ | V | 0.735 | 0.3273 | -3.493 | 0.775 | -5.29 |
| $V_{GS0L}$ | m | 7.96x10$^{-8}$ | 8.771x10$^{-7}$ | -1.67x10$^{-7}$ | -1.4x10$^{-8}$ | -1.02x10$^{-6}$ |
| $V_{GS0L2}$ | m$^2$ | -3.67x10$^{-15}$ | -6.396x10$^{-14}$ | -2.54x10$^{-15}$ | - | 2.31x10$^{-12}$ |
| delta | meV | 5.5 | 100.5 | 90 | 58 | 90 |
| deltaL | m | 8.95x10$^{-6}$ | -7.034x10$^{-8}$ | 3.31x10$^{-7}$ | 4.23x10$^{-7}$ | 2.88x10$^{-6}$ |
| deltaL2 | m$^2$ | -6.82x10$^{-13}$ | 2.402x10$^{-14}$ | -1.81x10$^{-14}$ | - | -1.35x10$^{-11}$ |
| $\theta_{int}$ | 1/V | 3.1 | 0.181 | 0.015 | 1.5 | 0.0258 |
| $u_{sat}$ | m/s | 9x10$^4$ | 2.5x10$^5$ | 9x10$^4$ | 9.273x10$^4$ | 3x10$^5$ |
| b | - | 2 | 2 | 2 | 2 | 2 |
| $K_{tr}$ | - | -0.4 | - | -1.2 | -0.777 | - |
| $\mu_{0p}$ | cm$^2$/(V·s) | 415 | 2419 | 928 | 938 | 2120 |
| $\mu_{pL}$ | m | -1.93x10$^{-7}$ | -1.423x10$^{-7}$ | -2.87x10$^{-7}$ | -1.49x10$^{-7}$ | -6.85x10$^{-7}$ |
| $\mu_{pL2}$ | m$^2$ | 2.31x10$^{-14}$ | 5.834x10$^{-15}$ | 2.08x10$^{-14}$ | - | -3.85x10$^{-12}$ |
| $\mu_{0n}$ | cm$^2$/(V·s) | 255 | 896.3 | 1340 | 938 | 1830 |
| $\mu_{nL}$ | m | -1.74x10$^{-7}$ | -1.269x10$^{-7}$ | -2.81x10$^{-7}$ | -1.49x10$^{-7}$ | -1.33x10$^{-7}$ |
| $\mu_{nL2}$ | m$^2$ | 1.72x10$^{-14}$ | 5.814x10$^{-15}$ | 2.12x10$^{-14}$ | - | -3x10$^{-12}$ |
| $R_{ds0p}$ | Ω | 108 | 82.3 | 76.8 | 51 | 20 |
| $R_{ds0n}$ | Ω | 108 | 133.4 | 229.3 | 51 | 43.5 |

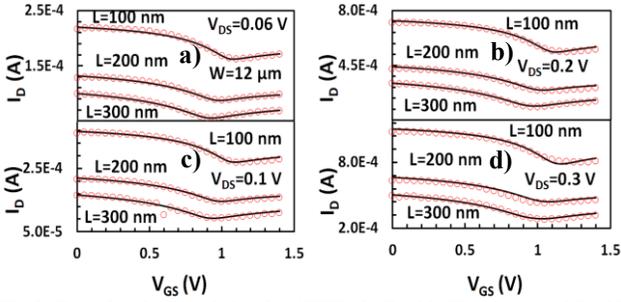

**Fig.3.** Transfer characteristics for GFETs [16] with *W=12 μm* and *L=100, 200, 300 nm* at $V_{DS}$*=0.06 V* (a), *0.1 V* (b), *0.2 V* (c) and *0.3 V* (d). Markers: measurements, lines: model.

Initially, the consistency of the updated scalable model vs. experiments for the Graphenea GFETs, is presented in the transfer characteristics of Fig. 2 for both *p*- and *n*-type regions; measurements are symbolized with markers and models with lines as in all the plots of the manuscript. Both measurements and models from the trap-reduced setups, are depicted in Fig. 2a-b for the *50/70, 100/50 μm/μm* GFETs, respectively, for every $V_{DS}$. Afterwards, remarkable model fittings with trap-affected experiments are shown in Fig. 2c-d for forward and backward sweeps (for the GFET of Fig. 2b), after the trap-related parameters [31] are extracted. Then, the coherence of the proposed geometrical scaling model (cf. (3)-(6)) is demonstrated vs. trap-reduced experiments for all GFETs, for $V_{DS}$*=0.1 V* (cf. Fig. 2e-f) as well as for $V_{DS}$*=0.7 V* (cf. Fig. 2g-h), confirming also its validity at both low and high $V_{DS}$ conditions.

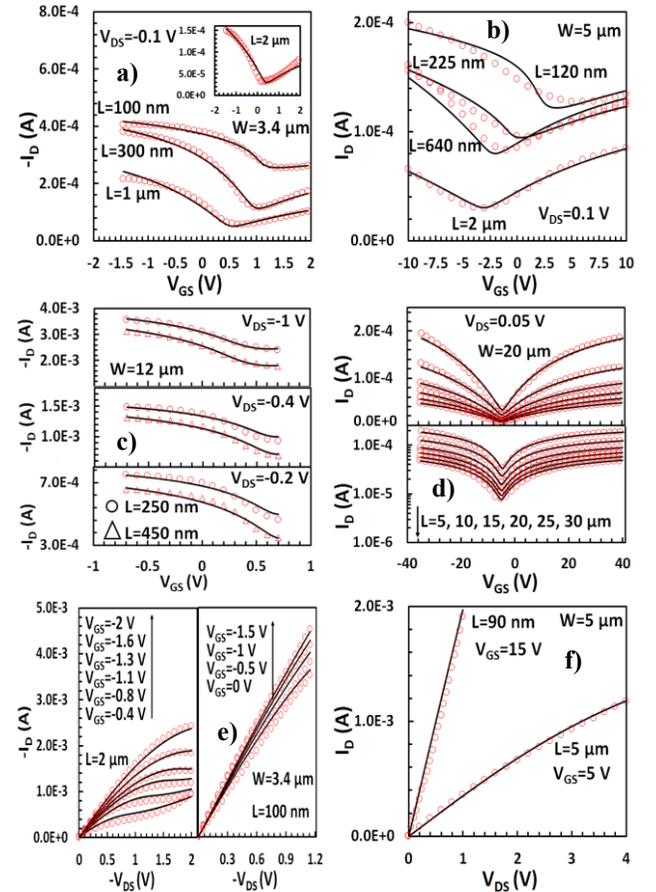

**Fig.4.** Transfer (a-d) and output (e-f) characteristics for GFETs with a, e) *W=3.4 μm* [17], b, f) *W=5 μm* [18], c) *W=12 μm* [19], d) *W=20 μm* in linear (upper panel) and logarithmic (bottom panel) y-axis [20] at $V_{DS}$*=-0.1 V* (a), *0.1 V* (b), *-0.2 V* (c-bottom panel), *-0.4 V* (c-center panel), *-1 V* (c-upper panel), *0.05 V* (d), $V_{GS}$=*-2, -1.6, -1.3, -1.1, -0.8, -0.4 V* (e-left panel), *-1.5, -1, -0.5, 0 V* (e-right panel) and *5, 15 V* (f) for long and short *L*. Markers: measurements, lines: model.

The model is also validated with measurements from GFET technologies obtained from literature for various short- [16]-[19] and long- [20] channel devices in terms of their transfer characteristics where, precise model vs. experiments' agreement is observed in every case. More specifically, three short-channel GFETs with $W=12\ \mu m$ and $L=300, 200, 100\ nm$ are demonstratred in Fig. 3a-d at $V_{DS}=0.06, 0.1, 0.2\ 0.3\ V$ [16]. Additionally, Fig. 4a introduces two relatively long- ($L=1, 2\ \mu m$) and two short- ($L=0.1, 0.3\ \mu m$) channel GFETs with $W=3.4\ \mu m$ and $V_{DS}=-0.1\ V$ [17]; $L=2\ \mu m$ device is shown in the inset of Fig. 4a. Similarly, Fig. 4b displays four GFETs with $W=5\ \mu m$ and $L=2, 0.64, 0.225$ and $0.1\ \mu m$, respectively, at $V_{DS}=0.1\ V$ [18]. In Fig. 4c, two short-channel GFETs ($L=0.25, 0.45\ \mu m$) are illustrated with $W=12\ \mu m$ and $V_{DS}=-0.2, -0.4, 1\ V$ in the bottom, middle and upper panels, respectively [19]. Six wide-long GFETs ($W=20\ \mu m$, $L=5, 10, 15, 20, 25, 30\ \mu m$) are also shown in Fig. 4d in logarithmic and linear y-axis in bottom and upper plots, respectively, for $V_{DS}=0.05\ V$ [20].

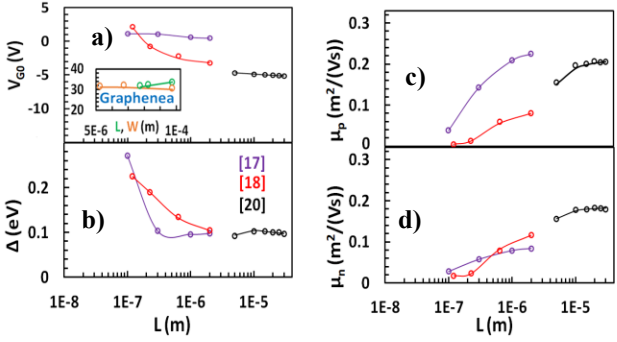

**Fig.5**. a) flat-band voltage $V_{G0}$, b) residual charge-related $\Delta$ and c, d) $p(n)$-type low-field mobility $\mu_{p(n)}$ parameters vs. $L$ (vs. $L$, $W$ in the inset) for GFETs from [17], [18], [20] (inset: Graphenea GFETs). Markers: extracted from measurements, lines: model derivations from (3), (5)-(6).

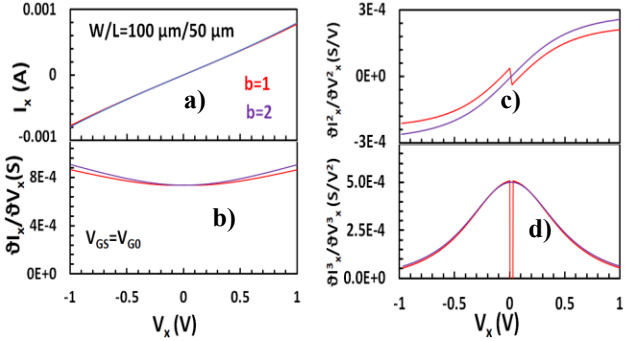

**Fig.6**. Gummel symmetry test: current $I_x$ (a), $\partial I_x/\partial V_x$ (b), $\partial I^2_x/\partial V^2_x$ (c) and $\partial I^3_x/\partial V^3_x$ (d) vs. channel potential $V_x$ for a GFET with $W=100\ \mu m$ and $L=50\ \mu m$ at $V_{GS}=V_{G0}$ (Dirac point) with velocity saturation model parameter $b=1, 2$ (cf. (2)).

To better validate the model at stronger horizontal electric fields, $V_{DS}$-dependent experiments from short-channel GFETs are essential. Fig. 4e-f depict the output characteristics of such short-channel GFETs from [17], [18], where again the model precisely fits the measurements. The devices with $L=2\ \mu m$, $V_{GS}=-0.4, -0.8, -1.1, -1.3, -1.6, -2\ V$ (left panel) and $L=100\ nm$, $V_{GS}=0, -0.5, -1, -1.5\ V$ (right panel) are presented in Fig. 4e [17] while those with $L=5\ \mu m$ ($V_{GS}=5\ V$) and $L=90\ nm$ ($V_{GS}=15\ V$) in Fig. 4f [18]. As $V_{DS}$ increases, VS and trapping effects become critical and the model accounts for these mechanisms through $u_{sat}$, $K_{tr}$ parameters [16], [31]. Both the above effects are notable in the devices from [16], [18], [19], while only VS seems to contribute in the device from [17] (cf. Table I). To investigate more thoroughly the geometrical dependence of those large longitudinal field-related parameters, additional measurements are required from more short devices, which is a task of a future work.

The $L$-dependence of $V_{G0}$, $\mu_{p(n)}$, $\Delta$ parameters is presented in Fig. 5a-d for GFETs from [17], [18], [20] ($W$, $L$-dependence of $V_{G0}$ for Graphenea GFETs is shown in inset of Fig. 5a) with remarkable agreement between experimental and model (cf. (3)-(6)) extractions. Note that $V_{G0}$ generally adds as $L$ decreases, especially at sub-micrometer range [17] but it can also slightly deviate for large $W$, $L$ (cf. Fig. 5a). The $\sim 1/L$ relation of $\mu_{(p)n}$ and $\sim L$ of $\Delta$ for shorter $L$ for all examined technologies are confirmed [26] (cf. Fig. 5b-d). The model interprets notably the measurements from strong $p$- to strong $n$-type region, including $V_{Dirac}$, and from low to high $V_{DS}$ with a negligible relative RMS error below 5% for most GFETs examined while it does not exceed 10% in any case (cf. Fig. 2-5).

A reliable benchmark for CMOS compact models to evaluate the symmetry and continuity around $V_{DS}=0\ V$, is the GST (cf. [2, Fig. K2]. To ensure accuracy, channel current must be an odd function of channel potential $V_x$ thus, current leakages can be avoided by probing the quantity $I_x=(I_D-I_S)/2$ [2, eq. K2] where $I_S$ is the source side current. In the special case of GFET, which is an inherently symmetrical device due to its ambipolarity, GST is principally examined at $V_{GS}=V_{G0}$, due to the drift of the Dirac point generated by gate doping [33]. Simulated $I_x$ and its 1st, 2nd and 3rd derivatives with $V_x$ are depicted vs. $V_x$ for a GFET with $W=100\ \mu m$ and $L=50$ in Fig.6a-d, respectively, for VS factor $b=1, 2$. The proposed model successfully passes GST with continuous and symmetrical results for $b=2$ for every case as cited in [2, §5.2], while unwanted singularities are noticed for higher order derivatives when $b=1$, as it is referred in [2, §5.2]. Similar results are obtained for other GFETs studied here.

## V. CONCLUSIONS

A scalable symmetric compact GFET model, that accurately describes geometrical effects, is proposed in the present study. It targets towards the design of the first GFET Process Design Kit, and eventually, to the transition of GFET technologies from research labs to the breakthrough of large-scale industry-based IC production. Polynomial and exponential equations are formulated through investigating geometry-induced dependencies of basic model parameters. The scalability features of the model presented here, precisely characterize experiments from different fabricated GFET technologies with different $W$ and from short ($nm$) to long ($\mu m$) gate devices. Furthermore, the unavoidable trap effects observed in this emerging technology are also captured by the model. The latter aids to select scenarios to enhance the reproducibility of the device characteristics.

## APPENDIX

$$Q_{gr} = Q_t + en_{res} = Q_p + Q_n + en_{res} = \frac{k}{2}V_c^2 + \frac{e\Delta^2}{\pi(hu_f)^2} \qquad (7)$$

where $Q_t$, $Q_{p(n)}$ are the transport sheet, $p(n)$-type charges, respectively; $V_c$ is the chemical potential, $h$ the reduced Planck constant, $u_f$ the Fermi velocity, $e$ the electron charge and $k$ a coefficient [5, §2.1]. $Q_t$, and thus $I_D$, can be calculated according to $V_c$ polarity at source ($V_{cs}$) and drain ($V_{cd}$), respectively. Hence, at $n$-type region where $V_{cs}$, $V_{cd}>0$, $Q_p=0$:

$$Q_t = Q_n = \frac{k}{2}V_c^2 \quad (8)$$

and thus, by using (7) and (8) and considering $dV/dV_c=-(C_q+C)/C$ [5, §2.1], [16], (1) becomes:

$$I_D = I_{Dn} + I_{Dp} = \frac{W\mu_{un}\int_{V_s}^{V_d}(|Q_n|+\frac{en_{res}}{2})dV}{L_{effn}} + \frac{W\mu_{up}\int_{V_s}^{V_d}(|Q_p|+\frac{en_{res}}{2})dV}{L_{effp}} =$$
$$\frac{W\mu_{un}\int_{V_s}^{V_d}(\frac{k}{2}V_c^2+\frac{en_{res}}{2})dV}{L_{effn}} + \frac{W\mu_{up}\int_{V_s}^{V_d}(0+\frac{en_{res}}{2})dV}{L_{effp}} = \frac{W\mu_{un}\int_{V_{cs}}^{V_{cd}}\frac{k}{2}V_c^2\frac{dV}{dV_c}dV_c}{L_{effn}} +$$
$$\frac{W\mu_{un}\int_{V_s}^{V_d}\frac{en_{res}}{2}dV}{L_{effn}} + \frac{W\mu_{up}\int_{V_s}^{V_d}\frac{en_{res}}{2}dV}{L_{effp}} = \frac{W\mu_{un}\int_{V_{cd}}^{V_{cs}}\frac{k}{2}V_c^2\frac{C_q+C}{C}dV_c}{L_{effn}} +$$
$$\frac{WV_{ds}en_{res}}{2}\left(\frac{\mu_{un}}{L_{effn}}+\frac{\mu_{up}}{L_{effp}}\right) \quad (9)$$

Similarly, at $p$-type region where $V_{cs}$, $V_{cd}<0$, $Q_n=0$:

$$Q_t = Q_p = \frac{k}{2}V_c^2 \quad (10)$$

and by using (7), (10), (1) yields:

$$I_D = I_{Dn} + I_{Dp} = \frac{W\mu_{up}\int_{V_{cd}}^{V_{cs}}\frac{k}{2}V_c^2\frac{C_q+C}{C}dV_c}{L_{effp}} + \frac{WV_{ds}en_{res}}{2}\left(\frac{\mu_{un}}{L_{effn}}+\frac{\mu_{up}}{L_{effp}}\right) \quad (11)$$

At ambipolar conduction where $V_{cs}>0$, $V_{cd}<0$:

$$I_D = I_{Dn} + I_{Dp} = W\mu_{un}\frac{\int_{V_s}^{V_z}(|Q_n|+\frac{en_{res}}{2})dV+\int_{V_z}^{V_d}(|Q_n|+\frac{en_{res}}{2})dV}{L_{effn}} +$$
$$W\mu_{up}\frac{\int_{V_s}^{V_z}(|Q_p|+\frac{en_{res}}{2})dV+\int_{V_z}^{V_d}(|Q_p|+\frac{en_{res}}{2})dV}{L_{effp}} =$$
$$W\mu_{un}\frac{\int_{V_s}^{V_z}(\frac{k}{2}V_c^2+\frac{en_{res}}{2})dV+\int_{V_z}^{V_d}(0+\frac{en_{res}}{2})dV}{L_{effn}} +$$
$$W\mu_{up}\frac{\int_{V_s}^{V_z}(0+\frac{en_{res}}{2})dV+\int_{V_z}^{V_d}(\frac{k}{2}V_c^2+\frac{en_{res}}{2})dV}{L_{effp}} = \frac{W\mu_{un}\int_0^{V_{cs}}\frac{k}{2}V_c^2\frac{C_q+C}{C}dV_c}{L_{effn}} +$$
$$\frac{W\mu_{up}\int_{V_{cd}}^{0}\frac{k}{2}V_c^2\frac{C_q+C}{C}dV_c}{L_{effp}} + \frac{WV_{ds}en_{res}}{2}\left(\frac{\mu_{un}}{L_{effn}}+\frac{\mu_{up}}{L_{effp}}\right) \quad (12)$$

$V_z$ is the channel potential where $V_c=0$ V and the integral for every case in (9), (11), (12) is solved from $V_{ca}$ to $V_{cb}$, as:

$$\int_{V_{ca}}^{V_{cb}}\frac{k}{2}V_c^2\frac{C_q+C}{C}dV_c = \frac{k}{2C}\int_{V_{ca}}^{V_{cb}}V_c^2\left(k\sqrt{V_c^2+C_1^2}+C\right)dV_c =$$
$$\frac{k}{2C}\left[\frac{1}{24}\left(8CV_c^3+3kV_c\sqrt{C_1^2+V_c^2}(C_1^2+2V_c^2)-3C_1^4k\ln[V_c+\sqrt{C_1^2+V_c^2}]\right)\right]_{V_{ca}}^{V_{cb}} \quad (13)$$

where $C_q=k\sqrt{V_c^2+C_1^2}$ is the quantum capacitance with $C_1=U_T\cdot\ln(4)$ [5, §2.1] and thermal voltage $U_T=k_BT/e=25.6$ mV at $300$ K [2, eq. 1.2.12] where $k_B$ is the Boltzmann constant and $T$ the temperature; $C=C_{top}+C_{back}$ where $C_{top(back)}$ is the top(back) oxide capacitance [5, §2.1]. (8)-(13) are derived for $V_{ds}>0$ but the model also accounts for $V_{ds}<0$ where $V_{cd}$, $V_{cs}$ polarities are opposite [16, §S.I. E]; $V_{cd}$, $V_{cs}$ are estimated by implementing a self-consistent Verilog-A algorithm [5, §2.1]. A phenomenological smooth function used for $R_{ds}$ at the ambipolar region, is given by:

$$R_{ds} = \frac{1}{4}\big[R_{dsp}[2+\tanh(-14V_{cs})+\tanh(-14V_{cd})] + R_{dsn}[2+\tanh(14V_{cs})+\tanh(14V_{cd})]\big] \quad (14)$$